\documentclass{article}
\usepackage[left=4cm,right=4cm]{geometry}
\usepackage{graphicx} % Required for inserting images
\usepackage{caption}
\usepackage{chemformula}
\usepackage{cite}
\usepackage{mathtools}
\usepackage{adjustbox}

\title{Tailoring the electronic properties of \ch{TiO2} monolayers for solar driven catalysis through transition metal doping}
\author{Kati Asikainen$^*$, Matti Alatalo, Marko Huttula and S. Assa Aravindh$^*$}
\date{%
    \normalsize\textit{Nano and Molecular Systems Research Unit, University of Oulu, FI-90014, Finland}\\
\begin{tabular}{l}
\\[2pt]
\small ${}^*$Corresponding authors: Kati.Asikainen@oulu.fi, Assa.SasikalaDevi@oulu.fi
\end{tabular}
}

\begin{document}

\maketitle
\begin{abstract}
Substitutional doping with transition metals is carried out in the Lepidocrocite phase - the stable monolayer geometry of \ch{TiO2}, using density functional theory (DFT) methods. The doping is carried out at the differently coordinated O atom cites, producing Janus monolayer geometries. Our results indicate that key fundamental properties for photocatalysis can be tuned via doping. Monolayers doped with Ag, Au, Pd and Pt are thermodynamically stable, amongst all considered doping possibilities, as evident from phonon band structure calculations. Electronic structure of the Janus monolayers alters significantly, compared to pristine \ch{TiO2}, owing to the emergence of mid-gap states. Reduced band gap arises from upward shift of the valence band,
suggesting enhanced visible-light response. Dopant atoms also introduce excess electrons in \ch{TiO2} monolayers, which are found to localize at a single Ti site. This induces ferromagnetism in the doped monolayers. Furthermore, charge separation between \ch{TiO2} and noble metal dopants is observed which is a key parameter in influencing the selectivity and activity of photocatalytic materials. Compared to the pristine \ch{TiO2} monolayer, the Janus structure can promote water adsorption, and the Janus monolayers exhibit significantly improved activity in the hydrogen evolution reaction. These findings suggest that engineering a novel Janus \ch{TiO2}-based monolayer with a noble metal layer on the other surface can offer a potential approach to improve photocatalytic performance over pristine \ch{TiO2}. \\[2pt]

\end{abstract}

\textbf{Keywords}: DFT, transition metal doping, Janus morphology, photocatalysis, \ch{TiO2} monolayers

\section{Introduction}
Two-dimensional (2D) materials with thicknesses of one or a few atomic layers have gained much interest since the discovery of graphene in 2004 \cite{graphene}. Compared to the respective bulk counterparts, the increased surface area and other associated artifacts, such as quantum confinement, gives rise to unique physicochemical properties in these materials. So far, various 2D materials have been reported beyond graphene including but not limited to transition metal dichalcogenides \cite{TMDs, J1, J2}, MXenes \cite{MXenes}, metal-organic frameworks \cite{MOF, J3} and metal oxides \cite{MOs}. Among these, metal oxides form an exciting class of 2D materials. The presence of oxygen atoms at the surface can alter surface energies and enable the tuning of material properties. For several decades, titanium dioxide (\ch{TiO2}) has been extensively studied for a variety of catalytic applications. Besides its bulk phases, \ch{TiO2} also exists in 2D crystal structures. Lepidocrocite monolayer of \ch{TiO2} is the only truly 2D phase of \ch{TiO2}, synthesized through exfoliation by means of soft-chemical procedures \cite{Sasaki-TiO2}. Thermodynamic stability of this material has further been established by phonon calculations \cite{phonon-TiO2}. Since the pioneering work of Fujishima and Honda \cite{FH}, \ch{TiO2} has been extensively studied for photocatalytic applications as the band alignment straddles both the oxidation and reduction potentials of water, \ch{TiO2} is suitable for overall water splitting. Similarly, 2D lepidocrocite \ch{TiO2} has been reported to have a proper band alignment for water splitting reactions and possess strong redox tendencies\cite{TiO2-OER, TiO2-band-gap}. The 2D structure enables electron confinement to the Ti 3d orbitals while holes are confined to the O 2p orbitals at the topmost surface \cite{TiO2-eh}, and this can  promote photocatalytic activity by suppressing the recombination of charge carriers.

Despite  several advantages, 2D \ch{TiO2} suffers from performance degradation, due to the larger band gap energy compared to it's 3D counterparts \cite{TiO2-band-gap}. When water splitting occurs in a defect-free monolayer, hole participation is required to initiate dissociation of \ch{H2O} and the first oxidation step of oxygen evolution reaction (OER) has a high probability for reverse process which can hinder the photocatalytic efficiency \cite{TiO2-OER}. In addition to this, hydrogen evolution reaction (HER) activity is also lowered in pure 2D \ch{TiO2} due to the lack of favorable reaction centers \cite{TiO2-H2}. Doping of \ch{TiO2} emerges as an effective approach to tune the material properties and improve photocatalytic performance in water splitting. In this scenario, noble metals are often incorporated into crystal structures, and numerous studies have demonstrated improvement in the photocatalytic performance as a consequence. In various systems, noble metal doping resulted in visible-light sensitization \cite{L1, L2, L3, L4, L5} which is an essential characteristic for photocatalytic materials. Furthermore, noble metals generally act as efficient electron traps, leading to  separation of photogenerated electrons and holes \cite{L1, L3, L6}. This suppresses the recombination rate, in correlation with enhanced photocatalytic activity. In photocatalytic applications, noble metals have shown superior performance in order to improve HER activity. Especially, Pt has been considered as one of the most promising elements to increase \ch{H2} production rates \cite{Pt1, Pt2}. In addition to Pt, other noble metals, such as Pd, Au and Rh, have been used to significantly improve the HER activity in \ch{TiO2}-based photocatalysts \cite{L6, HER1, HER2, HER3, HER4}. 

In \ch{TiO2} lattice a dopant can substitutionally replace either Ti or O atom. Replacement of Ti is generally well-studied but researchers have also reported successful replacement of oxygen sites. In \textit{et al.} \cite{Bo} have investigated boron doping in \ch{TiO2} anatase, and their results indicated preferred substitution of oxygen atoms by boron. More insight has been given by Patel \textit{et al.} \cite{Bo1} whose theoretical work suggested that boron preferentially replaces and occupies oxygen sites at high dopant concentration. Sulfur doping has also been studied extensively in \ch{TiO2} lattice. Umebayashi \textit{et al.} \cite{Su-O} suggested S anion doping on oxygen sites whereas Ohno \textit{et al.} \cite{Su-Ti} observed a substitution of Ti atoms by S cations. Similarly, C-doped \ch{TiO2} have been synthesized and replacement of lattice oxygen by carbon species was also reported \cite{Ca1, Ca2}. According to the DFT calculations conducted by Heffner \textit{et al.} \cite{C-TiO2}, substitional C doping on O site was more favorable than on Ti site in monoclinic \ch{TiO2} regardless of the growing conditions. Filippatos \textit{et al.} \cite{F-Cl} have prepared hydrogen, fluorine and chloride doped and codoped \ch{TiO2} in which F or Cl succesfully substituted a lattice oxygen. The studies have shown improved photocatalytic activity in the visible light region for \ch{TiO2} with doping on oxygen sites. Theoretical investigations have revealed an emergence of impurity states within the band gap which are proposed to be responsible for the red-shift in photoresponse \cite{Bo1, Su-O, Su-Ti, Su-O2, C-TiO2, F-Cl}. The studies demonstrate that substitutional doping on oxygen sites in \ch{TiO2} emerges as a potential method towards the development of visible light induced photocatalysts.

Previously, 2D \ch{TiO2} has been doped by substituting titanium atoms by metal atoms. Particularly, Cr and Mn-doping were demonstrated to optimize hydrogen adsorption to the corresponding level as Pt-doping, indicating substantially better HER activity compared with undoped \ch{TiO2} monolayer \cite{TiO2-HER}. Alternatively, Rh- and Pd-doping have been used to stabilize the rate limiting oxidation step of OER by reducing the overpotential and thermodynamic barriers \cite{TiO2-OER} and Rh-doping to increase the \ch{H2} production rate \cite{TiO2-H2}. However, substitution of  oxygen sites is not well explored in 2D \ch{TiO2}. Therefore, in this work, we investigate the incorporation of noble metal dopants as substitutional defects at oxygen sites of 2D lepidocrocite \ch{TiO2} using first principles calculations. A lepidocrocite monolayer consists of two types of oxygen: two-fold coordinated bridging atoms and four-fold coordinated in-layer atoms, providing two different doping sites for noble metal atoms. Presumably, a bridging oxygen atom is the most probable doping site, forming fewer chemical bonds in the monolayer. We perform a comprehensive investigation on formation, stability, and electronic properties of doped 2D \ch{TiO2} monolayer. It is found that monolayers doped with Ag, Au, Pd and Pt at bridging oxygen site are stable, and the doping leads to a formation of Janus structure in \ch{TiO2} monolayer. Analysis reveal that dopants alter the electronic structure, magnetism and charge carrier dynamics to a great extent. The present work provides an interesting strategy to modify the photocatalytic properties of \ch{TiO2} monolayer by creating a novel Janus structure via doping.

\section{Methodology}
The first principles calculations based on density functional theory were performed, as implemented in the Vienna Ab initio Simulation Package (VASP) \cite{vasp1, vasp2, vasp3}. We used the projector augmented wave (PAW) \cite{PAW} method to simulate the core-ion potential while valence electrons were represented using a plane wave basis set with an energy cutoff of 520\ eV. The generalized gradient approximation (GGA) with Perdew-Burke-Ernzerhof (PBE) functional was employed \cite{GGA}. Because of the localized d electrons of transition metals, we adopted the GGA+U formalism according to Dudarev \textit{et al.} \cite{GGA+U} for more accurate description of the investigated systems. Since in many cases, the GGA+U has been observed to overestimate the lattice parameters over the GGA \cite{TiO2-N/Ta, LiMn2O4}, we performed geometry optimization with the GGA and subsequently used the optimized GGA geometries in the GGA+U calculations. The  Brillouin zone was sampled according to the Monkhorst-Pack scheme \cite{MP} with a k-point grid of $6 \times 5 \times 1$ in the geometry optimization and $13 \times 10 \times 2$ in the electronic structure calculations. Gaussian smearing with a width of 0.05 eV was applied. The convergence criterion for energy and forces were set to $10^{-6}\ \mathrm{eV}$ and $0.001\ \mathrm{eV/\AA}$ in all calculations, respectively. The post-processing of the VASP outputs were done using the VASPKIT \cite{vaspkit} toolkit.

We considered a unit cell of 2D lepidocrocite \ch{TiO2} unit cell which comprises a total of six atoms: two Ti and four O atoms (Fig. \ref{2D-TiO2}a). To prevent interaction of periodically repeating images, a vacuum with a thickness of around 10 \AA\ was added along z-direction at both interfaces. As previously noted, two different types of oxygen atoms can be distinguished within the unit cell: bridging oxygen atom ($O_b$) and in-layer oxygen atom ($O_i)$, which are two-fold and four-fold coordinated (Fig. \ref{2D-TiO2}a), respectively. Conversely, all the titanium atoms are six-fold coordinated. Doped \ch{TiO2} monolayers were created by substituting either one $O_b$ or $O_i$ oxygen atom with a noble metal atom (Ag, Au, Pd, Pt, Rh, Ru, Ir or Os) within the unit cell. The particular doping concentration was employed in order to completely replace one of the atomic layers of \ch{TiO2} by noble metal atoms. Hence, replacement of $O_b$ could result in a Janus structure. The crystal structures of all systems were visualized using the VESTA \cite{VESTA} program. To assess the stability of the undoped and doped \ch{TiO2} monolayers, we used density functional perturbation theory to calculate the lattice dynamics. Followed by this, the PHONOPY \cite{phonopy} code was utilized to calculate the harmonic force constants and phonon band structures. The dipole corrections were not considered in the calculations except for the calculations of electrostatic potential, in order to  eliminate the artificial interactions caused by the periodic boundary conditions.

To study the catalytic properties of pristine and doped \ch{TiO2} monolayer, the optimized structure with $3 \times 3 \times 1$ supercell of the monolayers was taken, containing in total 54 atoms. Many different adsorption sites for the adsorbed molecules were examined, and the ionic relaxation was allowed during the geometry optimization. The most energetically stable adsorption structures were considered in this study. The same computational methodology was used except that the k-point grid was adjusted to $3 \times 2 \times 1$ in the calculations.

\section{Results}
\subsection{Optimization of the pristine \ch{TiO2} monolayer}
\begin{figure}[h!]\centering
\includegraphics[width=0.9\linewidth]{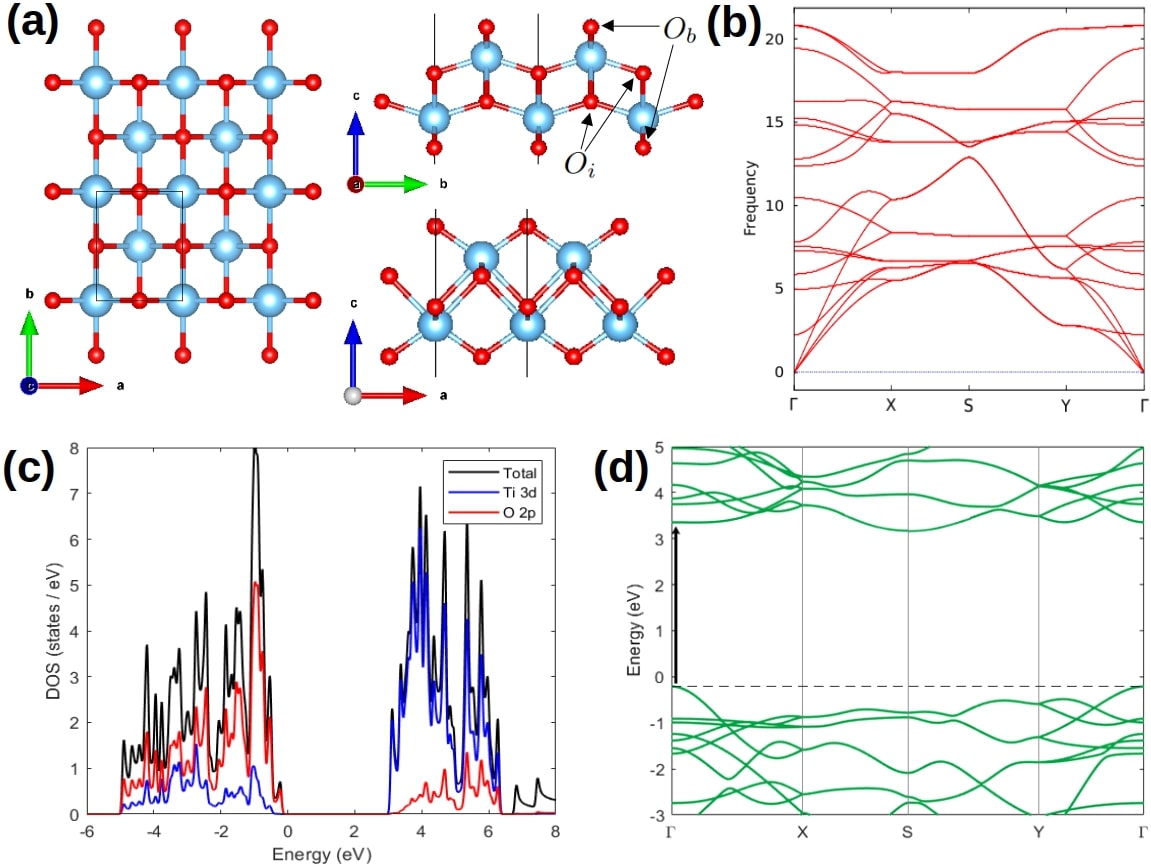}
\caption{(a) Schematic illustration of the structure of 2D lepidocrocite \ch{TiO2} monolayer. Ti atoms are marked as blue and O atoms as red here and elsewhere, and the unit cell is indicated with black lines. Furthermore, two different types of oxygen atoms, bridging and in-layer, are indicated with $O_b$ and $O_i$, respectively. (b) Calculated phonon band structure of the \ch{TiO2} monolayer. Electronic structure was investigated through the (c) Density of states (DOS) and (d) band structure with the Hubbard correction. The valence band maximum is marked with a vertical dashed line, and the direct band gap of 3.30 eV is indicated with a black line in (d).}\label{2D-TiO2}
\end{figure}

The crystal structure of 2D \ch{TiO2} monolayer is depicted in Fig. \ref{2D-TiO2}a. The optimized structural parameters of the \ch{TiO2} monolayer were $a=3.03\ \mathrm{\AA}$ and $b=3.77\ \mathrm{\AA}$, and the Ti-O bond lengths varied from 1.85 to 2.22\ \AA. These are consistent with previously reported experimental and theoretical values \cite{phonon-TiO2, P1,P3}. Fig. \ref{2D-TiO2}b shows the phonon band structure of the monolayer. No imaginary phonon modes were exhibited, thereby confirming the dynamical stability of the \ch{TiO2} monolayer. This is consistent with earlier evidence \cite{Phonon-TiO2}. The electronic structure analysis was carried out by calculating the density of states (DOS) and band structure using the GGA and GGA+U formalisms (Fig. S1 and Fig. \ref{2D-TiO2}). For Ti atoms a Hubbard correction of $U_{\mathrm{Ti}}=$4.5 eV was applied. As in bulk \ch{TiO2}, the DOS showed that the valence band of 2D \ch{TiO2} is mainly composed of O 2p states while the conduction band is dominated by Ti 3d states. Upon analysing the band structure, we found a direct band gap of 2.76 eV with the GGA (Fig. S1) and 3.30 eV with the GGA+U (Fig. \ref{2D-TiO2}c) located at $\Gamma$. The results underscore the inability of the GGA functional to describe the localized electrons, and applying the Hubbard correction greatly improves the description of \ch{TiO2} monolayer. The obtained band gap with the GGA+U aligns more closely with the experimentally measured optical band gap (3.8 eV) \cite{TiO2-band-gap}. Furthermore, the Bader charge analysis was performed, indicating that Ti lost 2.02\ \textit{e} in the \ch{TiO2}  monolayer, while the bridging and in-layer O atoms gain -0.91\ \textit{e} and -1.11\ \textit{e}, respectively. 

%lisää kommentit U(O) 

%muokkaa kuva!!!

\subsection{Noble metal doping of the \ch{TiO2} monolayer}

\subsubsection{Thermodynamic and energetic stability}
\begin{figure}[h!]\centering
\includegraphics[width=1\linewidth]{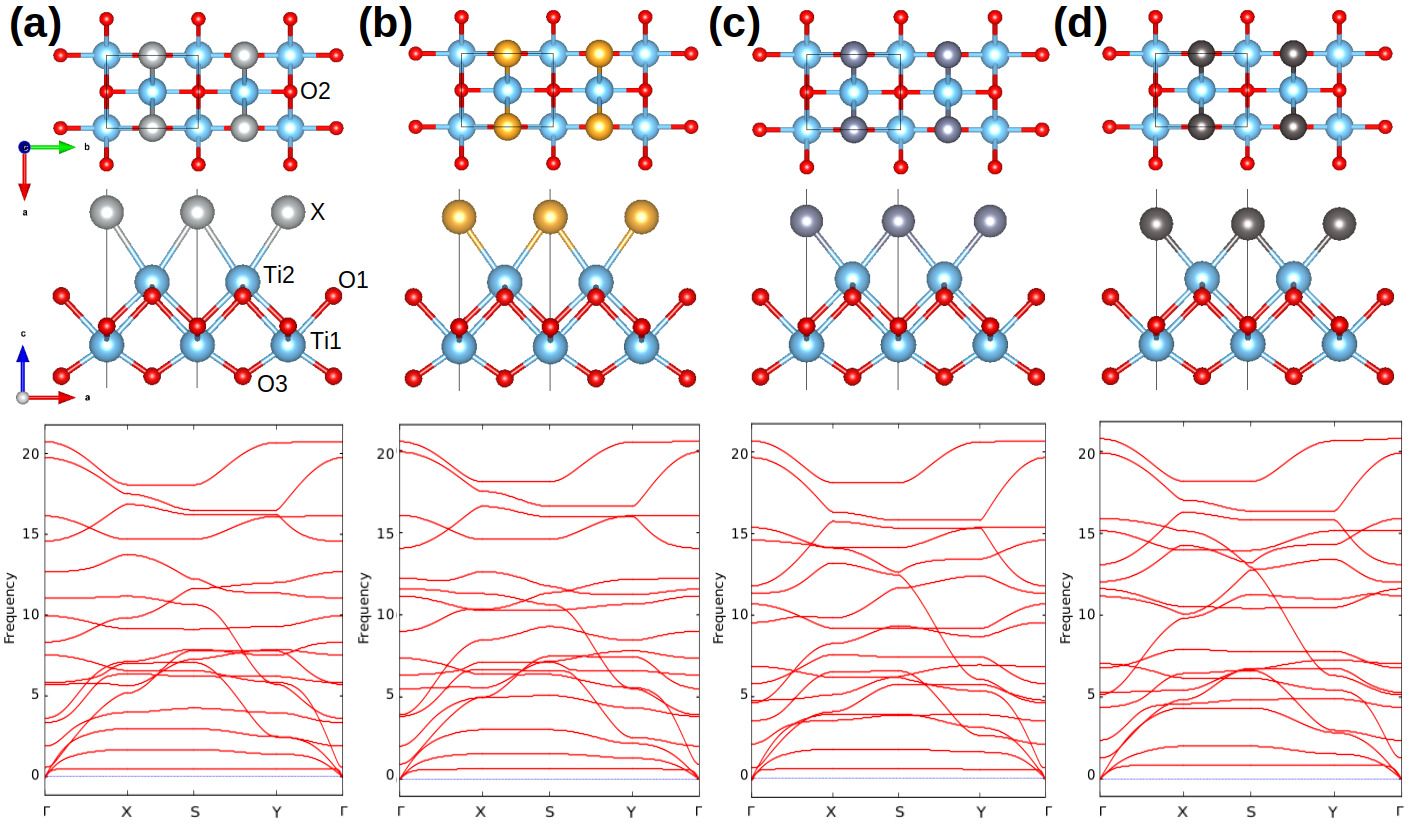}
\caption{Crystal structure and phonon band structure of the (a) Ag-\ch{TiO2}, (b) Au-\ch{TiO2}, (c) Pd-\ch{TiO2} and (d) Pt-\ch{TiO2} monolayers. Atom labeling is also shown in (a) where X specifies the noble metal dopant (X $=$ Ag, Au, Pd, Pt).}\label{Doped-MLs}
\end{figure}

We created 16 doped monolayers by substituting either $O_b$ or $O_i$ oxygen atom with a noble metal atom within the unit cell. After conducting geometry optimization, phonon dispersion calculations were performed to assess the dynamical stability of the systems. As a result, we found four stable \ch{TiO2} monolayers doped with Ag, Au, Pd and Pt on $O_b$ site. The monolayers were denoted as Ag-\ch{TiO2}, Au-\ch{TiO2}, Pd-\ch{TiO2} and Pt-\ch{TiO2}. Phonon band structures are depicted in Fig. \ref{Doped-MLs}, and the absence of imaginary phonon frequencies across the entire  Brillouin zone confirmed the dynamical stability of the structures. The remaining doped monolayers were found to be unstable, displaying imaginary phonon modes (Fig. S2 and S3). Therefore, we excluded them from the following discussion.

From now on, we only considered the four thermodynamically stable \ch{TiO2} monolayers doped with Ag, Au, Pd and Pt, and performed further analysis to gain insight into the properties of the systems. Fig. \ref{Doped-MLs} also represents the resulting crystal structures of the doped monolayers. The relaxed ground state structures were fairly similar compared to the undoped \ch{TiO2} monolayers. The substitution of a bridging oxygen atom led to the formation of a Janus structure for the doped monolayers, having distinct atomic layer on the opposite faces: noble metal atom and oxygen atom. The distances between the nearest Ti atom, labeled as Ti2, and the noble metal atom were 2.971\ \AA, 2.644\ \AA, 2.454\ \AA\ and 2.371\ \AA\ in the Ag-\ch{TiO2}, Au-\ch{TiO2}, Pd-\ch{TiO2} and Pt-\ch{TiO2}, respectively. As anticipated, the distance implicated the thickness of the monolayers. Given in the same order, the thicknesses were 5.531\ \AA, 5.356\ \AA. 5.187\ \AA\ and 5.074\ \AA. These are noticeably larger than that of undoped \ch{TiO2} monolayer, the thickness of which was 4.358\ \AA. The obtained thickness for the pristine \ch{TiO2} monolayer is consistent with other theoretical studies \cite{phonon-TiO2, P1}. In experiments, the thickness of synthesized monolayer has been reported to be around 7\ \AA\ or less than 10 \AA, in any case \cite{Phonon-TiO2, T1, T2, T3}. Further details of the crystal structure of the doped \ch{TiO2} monolayers are represented in Fig. S4. 
%figure: DOSs of doped monolayers GGA+U for Ti and NM

The dynamical stability was addressed through phonon calculations. Moreover, we studied energetic stability of noble metal defects by calculating the formation energy of the doped \ch{TiO2} monolayers using the following equation:
\begin{equation}
    E_{\mathrm{f}}=E(\mathrm{\ch{TiO2}+X})-E(\mathrm{\ch{TiO2}})+\frac{1}{2}E(\ch{O2})-E(\mathrm{X})
\end{equation}
where $E(\mathrm{\ch{TiO2}+X})$ and $E(\mathrm{\ch{TiO2}})$ are the total energies of noble metal doped and undoped \ch{TiO2} monolayers, respectively, where X specifies the noble metal atom (X $=$ Ag, Au, Pd, Pt). Energy of the Oxygen molecule is calculated by putting it in a cubic supercell, followed by relaxation. Finally, $E(\mathrm{X})$ is the single point energy of the noble metal atom, calculated from the bulk values. Our results showed that the defect formation in the monolayer is an endothermic process. The formation energy was calculated to be 9.80 eV in Ag-\ch{TiO2}, 8.37 eV in Au-\ch{TiO2}, 4.53 eV in Pd-\ch{TiO2} and 2.33 eV in Pt-\ch{TiO2}. It is worth noticing that Pd- and Pt-doping are more reasonable than Ag- and Au-doping, and Pt-doping is distinctly the most favorable, having over 4 times lower formation energy than Ag-doping. 

\subsubsection{Electronic structure analysis}
\begin{figure}[h!]\centering
\includegraphics[width=1\linewidth]{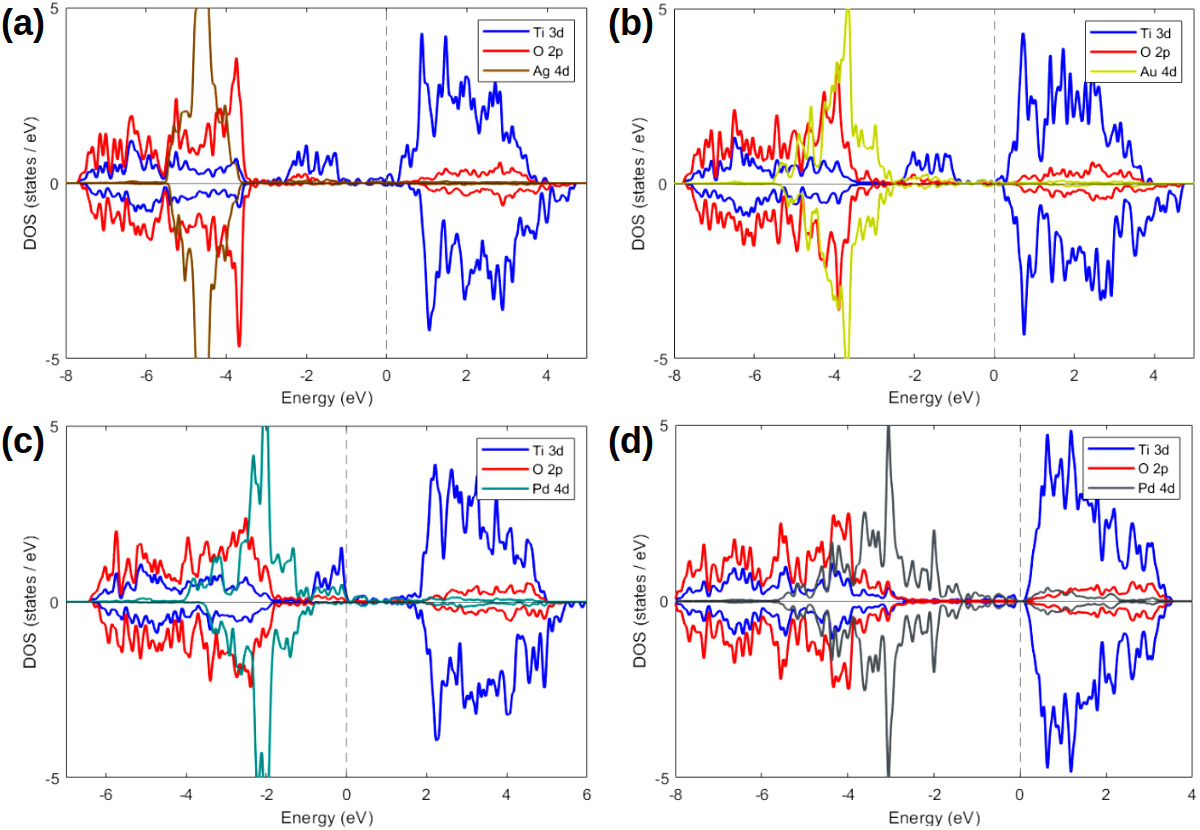}
\caption{Density of states of (a) Ag-\ch{TiO2}, (b) Au-\ch{TiO2}, (c) Pd-\ch{TiO2} and (d) Pt-\ch{TiO2} monolayers with the GGA+U. The Fermi level is set to zero and indicated with a black dashed line in the graphs.}\label{DOS}
\end{figure}

The effect of noble metal doping on the electronic structure of the Janus monolayers was investigated through the DOS and band structure. For each system, we calculated the DOS using both the GGA and GGA+U formalism to evaluate the performance of the Hubbard correction. To find reasonable Hubbard correction for Ag, Au, Pd and Pt, the linear response method \cite{LRM} was employed to estimate system-specific correction in the Janus monolayers. Based on our calculations, we selected the following $U$ corrections: $U_{\mathrm{Ag}}=$2.5 eV, $U_{\mathrm{Au}}=$3.0 eV, $U_{\mathrm{Pd}}=$2.0 eV, $U_{\mathrm{Pt}}=$2.5 eV. Fig. S5 and Fig. \ref{DOS} show the DOSs obtained with the GGA and GGA+U, respectively. In comparison to the traditional GGA, the Hubbard correction induced an opening of a band gap. This was attributed to the more accurate description of the localized valence electrons of metal atoms, causing the Ti and noble metal valence states to shift further away from the Fermi level. Moreover, the noble metal doping appeared to induce new Ti 3$d$ spin up defect states in Ag-\ch{TiO2}, Au-\ch{TiO2} and Pd-\ch{TiO2}. Using the GGA the newly emerging Ti 3d states were predicted to be delocalized and fused to the bottom of the conduction band whereas with the GGA+U, the states were well-localized within the band gap below the Fermi level. This is another significant difference in the performance of the formalism. Upon applying the Hubbard correction both opening the band gap and correcting the localization of defect states have been reported in several studies which have led to a better agreement of the theoretical and experimental results \cite{LRM, B1, B2, B3, B4}. Based on the observations, we concluded that the GGA+U improves the electronic structure description of the investigated Janus materials and therefore, we focused on analysing their properties with the GGA+U formalism.

\begin{figure}[h!]\centering
\includegraphics[width=1\linewidth]{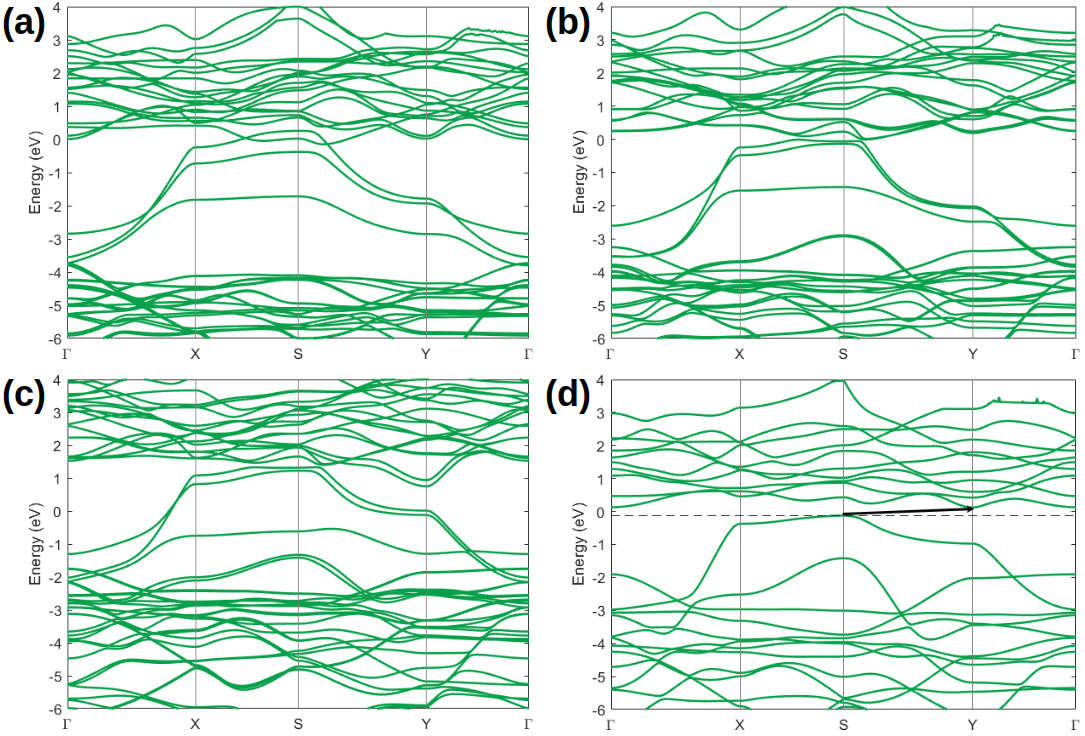}
\caption{Band structure of (a) Ag-\ch{TiO2}, (b) Au-\ch{TiO2}, (c) Pd-\ch{TiO2} and (d) Pt-\ch{TiO2} monolayers with the GGA+U. The valence band maximum and indirect band gap are indicated with a dashed vertical line and black arrow in (d), respectively. The Fermi level is set to zero.}\label{BS}
\end{figure}

Due to the incorporation of noble metal doping, noticeable changes were observed in the electronic structures. The doping led to an emergence of noble metal mid-gap states throughout the band gap of \ch{TiO2}, resulting in an upwards shift of the valence band  (Fig. \ref{DOS}). These can be beneficial to decrease the excitation energy of photogenerated electrons, enabling visible light adsorption, and to modify charge carrier dynamics \cite{L5, B4, mg1, mg2}. Interestingly, Ag-\ch{TiO2}, Au-\ch{TiO2} and Pd-\ch{TiO2} have characteristics of metallic behaviour because of some mid-gap states across the Fermi level. This was confirmed from the band structure plots (Fig. \ref{BS}). In particular, the localized hybridization of Ti and noble metal \textit{d}-states were observed at the Fermi level in the DOSs. In the Ag-\ch{TiO2} and Pd-\ch{TiO2}, the metallic character was evident, whereas in the Au-\ch{TiO2} the highest valence band and lowest conduction band barely overlapped, resulting in a zero band gap. This may contribute to an increase in catalytic activity, as shown in some studies \cite{m1, m2}. Pt-\ch{TiO2}, on the other hand, remained semiconducting under the noble metal doping, and the band gap was significantly decreased. We found an exceedingly small band gap of 0.25 eV (Fig \ref{BS}d). Moreover, because the valence band maximum located at S and conduction band minimum at Y, Pt-substitution led to a direct-to-indirect band gap transition in the \ch{TiO2} monolayer. Compared to the direct band gap, indirect band gap has a potential to prolong the existence of photogenerated electron-hole pair. Reported by Do \textit{et al.} \cite{MoSSe} and Behzad \cite{GeC} for instance, direct-to-indirect band gap transition has been realized previously in another Janus monolayer, MoSSe, and 2D GeC monolayer by introducing substitutional doping. In the Janus MoSSe monolayer, the doping was also reported to decrease the band gap energy substantially below 0.5 eV and aligning with our results that dopants can have a profound effect in 2D systems. %The Fermi level was also observed to significantly be shifted towards the conduction band in the doped monolayers. In the Ag-\ch{TiO2}, Au-\ch{TiO2} and Pt-\ch{TiO2} systems,  the Fermi level located right below the bottom of the conduction band of \ch{TiO2} while in the Pd-\ch{TiO2} it is located approximately in the middle of the band gap of \ch{TiO2}. 
Furthermore, the upward shift of the valence band edge indicated charge transfer occurring in these systems. We propose that the newly emerging Ti states that were well localized within the band gap correspond to excess electrons introduced by the noble metal dopants. The defect states were mainly associated with $e_g$ - orbitals, (Fig. S6), suggesting a predominant role for the orbitals in charge carrier dynamics.
%ks muistiinpanot 
% substantial number of emerging noble metal states may indicate electron trapping which can promote the separation of photogenerated electrons and holes in the monolayers. 

\subsubsection{Charge density analysis}
\begin{figure}[h!]\centering
\includegraphics[width=0.85\linewidth]{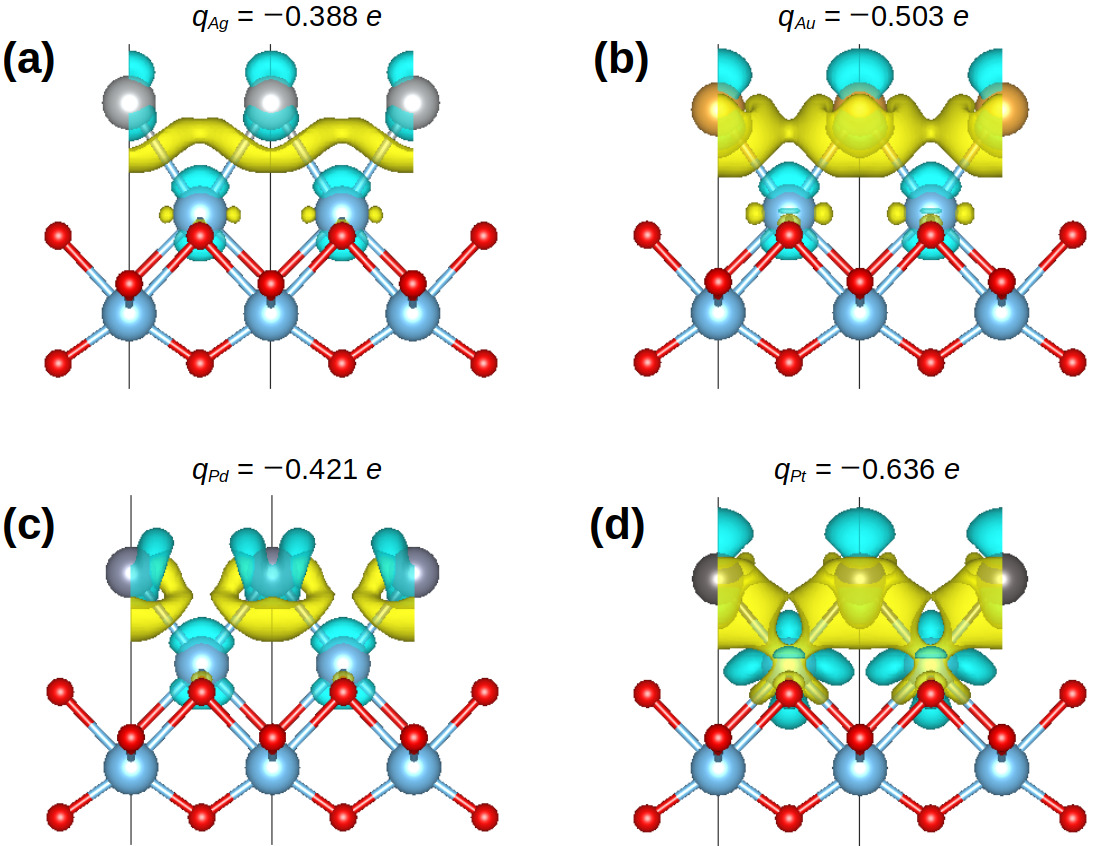}
\caption{Charge density difference after substitution of noble metal atom of (a) Ag-doped, (b) Au-doped, (c) Pd-doped and (d) Pt-doped  \ch{TiO2} obtained within the GGA+U formalism. Isosurface is set to 0.004\ \textit{e}/\AA³. Bader charge \textit{q} of the noble metal atom is given for each system. A negative Bader charge confirms the electron gain of noble metal atoms.}\label{CDD}
\end{figure}

To conduct charge density analysis, we calculated the charge density difference according to
\begin{equation}
    \Delta \rho = \rho(\mathrm{\ch{TiO2}-X})-\rho(\ch{TiO2})-\rho(\mathrm{X}).
\end{equation}
Here, $\rho(\ch{TiO2}+X)$, $\rho(\ch{TiO2})$ and $\rho(\mathrm{X})$ are the charge densities of the doped \ch{TiO2} monolayer, oxygen reduced \ch{TiO2} monolayer and noble metal atom, respectively. Moreover, we performed the Bader charge analysis \cite{Bader1, Bader2} to quantitatively investigate the charge redistribution. The charge density difference plots with the GGA+U are shown in Fig. \ref{CDD} in which yellow isosurface refers to charge accumulation and turquoise to charge depletion. In each doped monolayer the charge redistribution was localized at the interface of \ch{TiO2} and the dopant. The charge accumulation was observed in the bond regions between the dopant and Ti2 whereas anti-bonding states were concentrated near the dopant and Ti2 atom. The results suggest a tendency for the noble metal dopant and the Ti2 atom to form a chemical bond, which results in Janus monolayer geometries. 

The calculated Bader charges $q$ are listed in Table \ref{Bader}, where a negative Bader charge corresponds to electron gain and positive Bader charge to electron loss. The analysis confirmed electron transfer from \ch{TiO2} to the noble metal atom in each system, indicating ionic bonding between the noble metal dopant and the Ti2 atom. In ascending order, we found an electron transfer of 0.388\ \textit{e}, 0.421\ \textit{e}, 0.503\ \textit{e} and 0.636 \textit{e} per unit cell from \ch{TiO2} to the noble metal in Ag-\ch{TiO2}, Pd-\ch{TiO2}, Au-\ch{TiO2} and Pt-\ch{TiO2}, respectively (Fig. \ref{CDD}). The results indicate an improved spatial separation of electrons and holes in the doped \ch{TiO2} monolayer. This can be attributed to the Janus structure, which is capable of separating the charge carriers into different surfaces. This can lead to decreased recombination rate and increase in photocatalytic activity \cite{Nolan, MoSSe}. The greater electron transfer was reflected in the charge density difference plots, resulting in a more substantial charge redistribution at the interface of the dopant and Ti2. Furthermore, the localized defect states observed in the DOSs were confirmed to correspond to excess electron charge that is strongly localized at the Ti2 site. Due to the localization of charge, the Bader charge of the Ti2 decreased from 2.02\ \textit{e} to 1.61-1.69\ \textit{e} in the Ag-\ch{TiO2}, Au-\ch{TiO2} and Pd-\ch{TiO2} which, according to Posysaev \textit{et al.} \cite{Bader-TiO2}, suggests a change in the oxidation state of Ti from Ti$^{4+}$ to Ti$^{3+}$. 
\begin{table}[h]\centering
\begin{tabular}{llclclclclc}
\textbf{System} & & \multicolumn{1}{l}{\textbf{\ch{TiO2}}} & & \multicolumn{1}{l}{\textbf{Ag-\ch{TiO2}}} &  & \multicolumn{1}{l}{\textbf{Au-\ch{TiO2}}} &  & \multicolumn{1}{l}{\textbf{Pd-\ch{TiO2}}} &  & \multicolumn{1}{l}{\textbf{Pt-\ch{TiO2}}} \\ \hline 
\rule{0pt}{4ex}Ti1 & & 2.02  & & 2.00  &  & 2.01  &  & 2.02  & & 2.02  \\
\rule{0pt}{4ex}Ti2 & & 2.02  &  & 1.61  &  & 1.69  &  & 1.61 & & 1.78  \\
\rule{0pt}{4ex}O1 & & -1.11  &  & -1.17  &  & -1.15 &  & -1.17 & & -1.13  \\
\rule{0pt}{4ex}O2 & & -1.11  &  & -1.12  &  & -1.13 &  & -1.13 & & -1.12  \\
\rule{0pt}{4ex}O3 & & -0.91  &  & -0.923 &  & -0.917  &  & -0.913 & & -0.916  \\
\rule{0pt}{4ex}X  & & -0.91  &  & -0.388 &  & -0.503  &  & -0.421 & &  -0.636  \\[0.2cm] \hline                  
\end{tabular}
\caption{Bader charges (in \textit{e}) in \ch{TiO2}, Ag-\ch{TiO2}, Au-\ch{TiO2}, Pd-\ch{TiO2} and Pt-\ch{TiO2} with the GGA+U. In the undoped \ch{TiO2}, X corresponds to the fourth oxygen atom of the unit cell. In the doped systems, X specifies the noble metal dopant (X$=$ Ag, Au, Pd). Atom labeling is shown in Fig. \ref{Doped-MLs}(a).}\label{Bader}
\end{table}

\begin{figure}[h!]\centering
\includegraphics[width=1\linewidth]{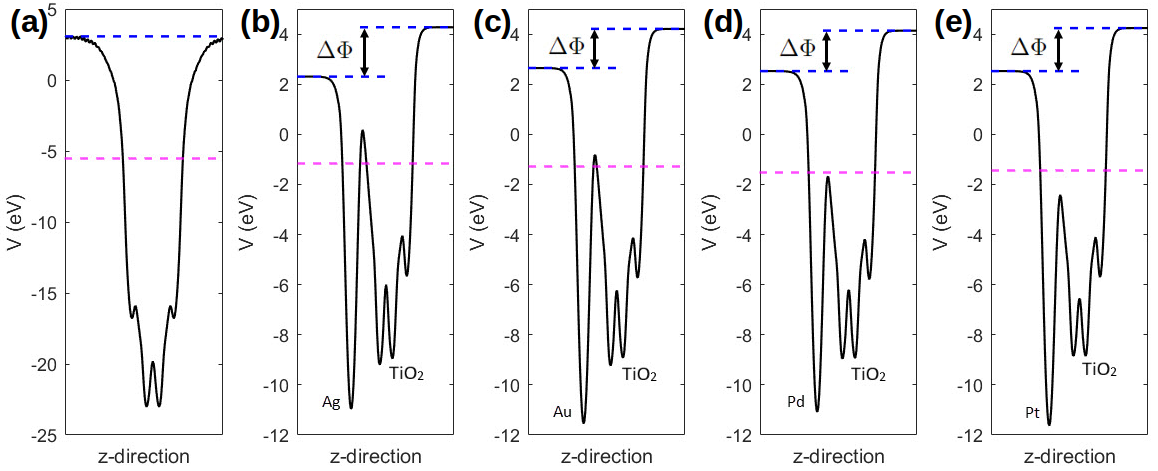}
\caption{Electrostatic potential of (a) undoped \ch{TiO2} monolayer, (b) Ag-\ch{TiO2}, (c) Au-\ch{TiO2}, (d) Pd-\ch{TiO2} and (e) Pt-\ch{TiO2}. The vacuum levels of the two surfaces are indicated with blue dashed lines, and the Fermi level with a purple dashed line. The electrostatic potential difference is represented by $\Delta \Phi$.}\label{WF}
\end{figure}

In Janus monolayers, the lack of out-of-plane inversion symmetry leads to uneven charge distribution between its surfaces and thus the existence of intrinsic electric field and permanent dipole moment. To further investigate the charge distribution we calculated the electrostatic potential of the undoped \ch{TiO2} and Janus monolayers, and the work function $\Phi$ by subtracting the Fermi level $E_F$ from the vacuum level $E_{\mathrm{vacuum}}$, 
\begin{equation}
\Phi = E_F-E_{\mathrm{vacuum}}.    
\end{equation} 
Due to the supercell approximation, both the asymmetric slab and the existence of the dipole moment can contribute to introducing artifacts into the calculations \cite{vacuum-potential}. Therefore, it is unlikely to obtain a flat vacuum potential in such Janus materials. However, this spurious interaction between the periodic images can be eliminated by applying a dipole correction method which will flatten the electrostatic potential in vacuum \cite{vacuum-potential,DC1,DC2}. We calculated the electrostatic potential of Janus monolayers with and without the dipole correction, and the results are shown in Fig. \ref{WF} and S7. As can be seen from Fig. S7, without the dipole correction the potential has a slope in the vacuum, due to which the vacuum level cannot be accurately determined. Therefore, it is crucial to apply the dipole correction to be able to calculate the work function with high accuracy. We found a large work function of 8.60 eV for the undoped \ch{TiO2} monolayer, as shown in Fig. \ref{WF}a. The doping was observed to considerably decrease the value, which can facilitate electron injection from the surface. Moreover, due to the Janus morphology the doped \ch{TiO2} monolayers exhibited dissimilar work function in the noble metal and oxygen surfaces, which were as follows: 3.47 eV and 5.44 eV in Ag-\ch{TiO2}, 3.84 eV and 5.50 eV in Au-\ch{TiO2}, 4.06 eV and 5.67 eV in Pd-\ch{TiO2}, and 3.98 eV and 5.69 eV in the Pt-\ch{TiO2}, respectively. This leads to an electrostatic potential difference $\Delta \Phi$ between the two surfaces with the values of 1.97 eV, 1.66 eV, 1.61 eV and 1.71 eV. This suggests that the noble metal substitution creates an intrinsic electric field along the out-of-plane direction, and the doped monolayers possess an intrinsic dipole \cite{D1, D2, D3}. Because the noble metal surface has a lower work function, the electric field is pointing from the noble metal to \ch{TiO2}. This is opposite to the electron transfer direction, which suggests that the intrinsic electric field can hinder the charge separation in the doped monolayers. The Ag-\ch{TiO2} possessed the lowest work function at both surfaces. On the other hand, Au and Pd-doping have the most favorable influence on the electron transfer from \ch{TiO2} to the dopants, resulting in lowest electrostatic potential differences. In Ag-\ch{TiO2} the largest electrostatic potential difference may be correlated with lowered local charge distribution and less efficient charge separation.

\subsubsection{Magnetic moments}
Based on the asymmetric DOSs, the Ag-\ch{TiO2}, Au-\ch{TiO2} and Pd-\ch{TiO2} appeared to show spin polarization due to the doping. After analysing the magnetism, the particular dopants (Ag, Au and Pd) indeed turned out to induce ferromagnetism in non-magnetic \ch{TiO2}. Simultaneously, the Pt-\ch{TiO2}, having symmetric spin up and down channels, was confirmed to be a non-magnetic system. The magnetic moments in the Ag-\ch{TiO2}, Au-\ch{TiO2} and Pd-\ch{TiO2} are shown in Table \ref{magmoms}. According to our findings, the magnetism originated mostly from the Ti2 atom which was also the origin of the localized excess charge states in the DOS. In each system, the Ti2 have a local magnetic moment of around 1\ $\mathrm{\mu_B}$, generally described as Ti$^{3+}$ \cite{B4, LM}. Ag-doping also induced moderate spin magnetization at Ti1 site, and in addition, magnetic moments of O atoms were somewhat larger compared to the values in the Au-\ch{TiO2} and Pd-\ch{TiO2}. This suggests that the charge is not as strongly localized and that the charge redistribution is extended to a wider area in the monolayer. The Ag and Au-dopant gained a very small amount of magnetization but in the Pd-\ch{TiO2} the magnetic moment of the Ti2 and Pt changed significantly in the system within the GGA+U formalism, thus representing an interesting case. For comparison, the magnetic moments calculated with the GGA are listed in Tab. S1. It was remarked that when the Hubbard correction was used, the magnetic moment of the Ti2 atom increased around 0.200\ $\mathrm{\mu_B}$ in the Ag-\ch{TiO2} and Au-\ch{TiO2} while in the Pd-\ch{TiO2} it increased over 0.7\ \ $\mathrm{\mu_B}$ compared to the value predicted by the GGA. Similarly, the magnetic moment of Pd was strongly affected by the Hubbard correction. Pd has a large Stoner factor of 0.874 \cite{Pd-Stoner} which is close to the Stoner criterion \cite{Stoner}. Bulk Pd as well as an isolated Pd atom are non-magnetic but it has been shown that a specific environment, such as low-dimensionality \cite{Pd-M1, Pd-M3}, can invoke magnetism. Being close to the Stoner transition point, Pd can be sensitive to on-site Coulomb interaction or any changes in the atomic structure, and this may explain the difference in magnetism with the GGA and GGA+U. For instance, in the study of Sirajuddeen \textit{et al.} \cite{PdO}, no spin polarization was predicted in PdO by the GGA, whereas with the GGA+U ferromagnetic behaviour originates in the system.
\begin{table}[h]\centering
\begin{tabular}{llclclc}
\textbf{System} &  & \multicolumn{1}{l}{\textbf{Ag-\ch{TiO2}}} &  & \multicolumn{1}{l}{\textbf{Au-\ch{TiO2}}} &  & \multicolumn{1}{l}{\textbf{Pd-\ch{TiO2}}} \\ \hline 
\rule{0pt}{4ex}Ti1                  &  & 0.135                       &  & 0.007                       &  & 0.004                       \\
\rule{0pt}{4ex}Ti2                  &  & 1.092                       &  & 1.00                        &  & 0.954                       \\
\rule{0pt}{4ex}O1                   &  & -0.026                      &  & -0.021                      &  & -0.020                      \\
\rule{0pt}{4ex}O2                   &  & -0.016                      &  & -0.002                      &  & 0.008                       \\
\rule{0pt}{4ex}O3                   &  & -0.020                      &  & -0.001                      &  & 0.001                       \\
\rule{0pt}{4ex}X                  &  & 0.015                       &  & 0.007                       &  & 0.074                       \\ 
\rule{0pt}{4ex}Total                &  & 1.18                        &  & 0.991                       &  & 1.021   \\[0.2cm] \hline                  
\end{tabular}
\caption{Magnetic moments (in $\mathrm{\mu_B}$ in Ag-\ch{TiO2}, Au-\ch{TiO2} and Pd-\ch{TiO2} in the unit cell. Total magnetic moment, the sum of local magnetic moments of individual atoms, of the unit cell is also provided. X specifies the noble metal dopant (X$=$ Ag, Au, Pd). Atom labeling is shown in Fig. \ref{Doped-MLs}(a).}\label{magmoms}
\end{table}

\subsection{Catalytic properties of pristine and doped \ch{TiO2} monolayer}
After examining the fundamental properties of the monolayers for photocatalysis, we investigated the performance of the Janus monolayers in photocatalysis through chemical reactions. We started by adsorbing a water molecule on the monolayers, followed by the evaluation of the HER performance.

\subsubsection{Adsorption of water molecule}
Adsorption of water is a crucial step to initiate a photocatalytic reaction of water splitting, and it can be used to characterize the activity of the material in the water splitting process. We explored the interaction between the investigated monolayers and absorbed a water molecule by calculating the adsorption energy from the following equation
\begin{equation}
E_{\mathrm{ads}}=E_{\mathrm{monolayer+\ch{H2O}}}+E_{\mathrm{monolayer}}+E(\ch{H2O}),
\end{equation}
where $E_{\mathrm{monolayer+\ch{H2O}}}$, $E_{\mathrm{monolayer}}$ and $E(\ch{H2O})$ are the total energy of the \ch{H2O} adsorbed on a monolayer, pristine monolayer and isolated water molecule. On the pristine \ch{TiO2} surface the water molecule is found to adsorb on top of the in-layer oxygen site with the O atom pointing to the surface (Fig. S8a). The charge was observed to accumulate between the \ch{H2O} molecule and the nearest Ti atom (Fig. S8b). The distance between these atoms was 2.57 \AA. The calculated adsorption energy was -0.25 eV, suggesting weak van der Waals interaction between \ch{H2O} and \ch{TiO2}. This value is consistent with the value reported by Casarin \textit{et al.} \cite{Casarin}, and similar to that of the value of \ch{H2O} on other 2D materials, such as MoSSe \cite{MoSSe-H2O} and \ch{WSSe} \cite{WSSe-H2O} monolayers.

Fig. S9 and S10 show the optimized adsorption structures with a \ch{H2O} molecule on the oxygen and noble metal surfaces of the Janus monolayers. At the oxygen surface, the $E_{\mathrm{ads}}$ value was -0.26 eV in Ag-\ch{TiO2}, -0.27 eV in Au-\ch{TiO2}, -0.32 eV Pd-\ch{TiO2} and -0.27 eV in Pt-\ch{TiO2}. The energies are 0.01-0.07 eV more negative than that of the value of pristine \ch{TiO2}, indicating slightly stronger interaction with a water molecule. However, the results don't suggest significant differences in adsorption behavior compared to the pristine \ch{TiO2}. The shortest distance between \ch{H2O} and surface oxygen varied in the range of 2.07 to 2.27 \AA\ (Fig. S9). Instead, at the noble metal surface the interaction was evidently strengthened, indicating more favorable binding with \ch{H2O}. The adsorption energy of \ch{H2O} decreased to -1.42 eV, -0.38 eV, -2.39 eV and -1.59 eV in Ag-\ch{TiO2}, Au-\ch{TiO2}, Pd-\ch{TiO2} and Pt-\ch{TiO2}, respectively. The trend of adsorption strength is thus Pd-\ch{TiO2} $>$ Pt-\ch{TiO2} $>$ Ag-\ch{TiO2} $>$ Au-\ch{TiO2}. The charge depletion region was observed between the oxygen atom of \ch{H2O} and the nearest noble metal atom, concentrating closer to the noble metal atom, while some electron transfer between the nearest hydrogen and noble metal atom might also occur (Fig. S10). The calculated DOSs (Fig. S11) revealed induced \ch{H2O} states in the valence band far below the Fermi level which overlap with the noble metal states to some extent. This indicates interaction between the noble metal dopant and water that could be identified as physisorption. The shortest distance between \ch{H2O} and the nearest noble metal atom was in the range of 2.49 to 2.92 \AA\ (Fig. S10). Upon adsorption both on the oxygen and noble metal surface, the H-O-H angle was either contracted or enlarged and the H-O bonds were altered compared to an isolated \ch{H2O} molecule (Table S2). One of the two H-O bonds was clearly elongated which could imply a weakened bond. This might be beneficial for the dissociation of \ch{H2O} on the surface \cite{H2O}. Therefore, the results show the Janus morphology can significantly alter and promote the adsorption of water and thus, have a positive impact on photocatalytic water splitting \cite{MoSSe-H2O, SnS2, MoS2-H2O}. 

\subsubsection{HER activity}
The Janus structures showed enhanced ability to capture \ch{H2O} on the noble metal surface. Because the charge density analysis indicated an electron transfer to the noble metal surface after the formation of the Janus monolayers, we investigated the hydrogen adsorption on the noble metal layer of the monolayers to evaluate their suitability for photocatalytic hydrogen production. We started by calculating the adsorption energy of hydrogen on the surface of the monolayers as follows 
\begin{equation}
E_{\mathrm{ads}} = E_{\mathrm{monolayer+\ch{H}}}+E_{\mathrm{monolayer}}+\frac{1}{2}E(\ch{H2}), 
\end{equation}
where $E_{\mathrm{monolayer+H}}$ and $E_{\mathrm{monolayer}}$ are the total energy of the monolayers with and without hydrogen adsorption, and the energy of an isolated hydrogen atom is calculated by dividing the total energy of the \ch{H2} molecule by two. The total energy of the \ch{H2} molecule is calculated by the cubic supercell box approach. To assess the HER activity of the materials, we calculated the Gibbs free energy ($\Delta G$) of hydrogen adsorption according to
\begin{equation}
\Delta G=E_{\mathrm{ads}}+\Delta E_{\mathrm{ZPE}}-T\Delta S,
\end{equation}
where $\Delta E_{\mathrm{ZPE}}$ and $T\Delta S$ represent the zero-point energy difference and entropy difference between the adsorbed hydrogen and hydrogen in a gas phase. Details of the calculations are given in the supplementary material. A catalyst with $\Delta G$ value of zero can be considered as an ideal catalyst for HER. The most stable adsorption structures are shown in Fig. S12.
\begin{figure}[h!]\centering
\includegraphics[width=0.8\linewidth]{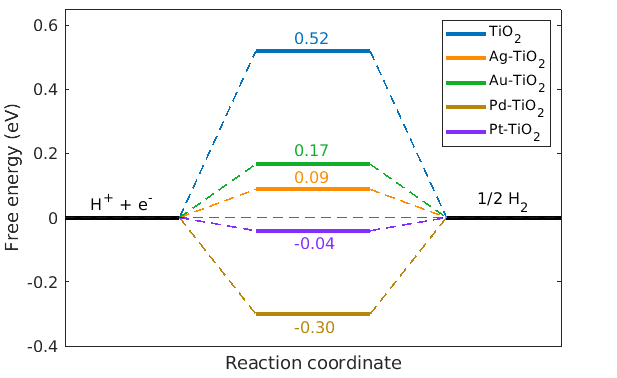}
\caption{The adsorption free energy diagram for HER on pristine \ch{TiO2} and Janus monolayer.}\label{HER}
\end{figure}

The calculated free energy diagram of HER is depicted in Fig. \ref{HER}. The Gibbs free energy of the pristine \ch{TiO2} monolayer was calculated to be 0.52 eV which is far from the ideal value and  the weak binding hinders the efficiency of the HER process. Previously, using a different computational setup, Yuan \textit{et al.} \cite{TiO2-HER} have reported a higher $\Delta G$ value of 1.09 eV for 2D lepidocrocite \ch{TiO2}. The formation of a Janus structure was observed to affect the adsorption behavior of hydrogen and modulated the adsorption free energy. The corresponding $\Delta G$ values of the Ag-\ch{TiO2}, Au-\ch{TiO2} and Pt-\ch{TiO2} were 0.09 eV, 0.17 eV, and -0.04 eV respectively, much lower than that of pristine \ch{TiO2}. Pt-doped \ch{TiO2} exhibited the lowest  $\Delta G$, and HER performance of both Ag-\ch{TiO2} and Pt-\ch{TiO2} are competitive to that of the Pt(111) surface (-0.09 eV) \cite{Pt}. For hydrogen adsorbed on Pd-\ch{TiO2}, $\Delta G$ had a larger negative value of -0.30 eV. This indicates too strong interaction with hydrogen which inhibits the desorption of hydrogen and the formation of \ch{H2} molecules. The results show that the Janus monolayers exhibit a significantly improved HER performance compared to pristine \ch{TiO2} monolayer, demonstrating that the HER activity can be optimized by engineering the Janus geometry characteristics.

\section{Conclusions}
2D \ch{TiO2} monolayer were engineered for better catalytic performance, by substituting O atoms by noble metals to produce Janus morphology. Phonon analysis demonstrated that monolayers doped with Ag, Au, Pd and Pt at two-fold coordinated bridging oxygen sites were thermodynamically stable. The presence of metallic atoms instead of O, induced additional electronic states and promoted charge kinetics, altering the electronic and magnetic properties of the Janus monolayers. Thus, these doped 2D Janus monolayers can effectively facilitate electron excitation under visible light and separate photogenerated charge carriers compared to uniform 2D materials, leading to improved photocatalytic performance. Investigations of chemical reactions indicated that the Janus structures resulted in a more favorable binding with a water molecule, and exhibited an increased HER activity. In general, interesting results from our investigation can pave way for future experiments on transition metal doped Janus monolayer systems. 

\section*{Acknowledgement}
This work has received funding from the European Union’s Horizon Europe research and innovation program under the Marie Sk\l odowska-Curie grant agreement no. 101081280. Further, CSC - IT Center for Science, Finland, is acknowledged for computational resources.

\section*{Declaration of competing interest}
The authors declare no competing financial interests.

\end{document}

\maketitle
\vspace{2cm}

\begin{figure}[h!]\centering
\includegraphics[width=1\linewidth]{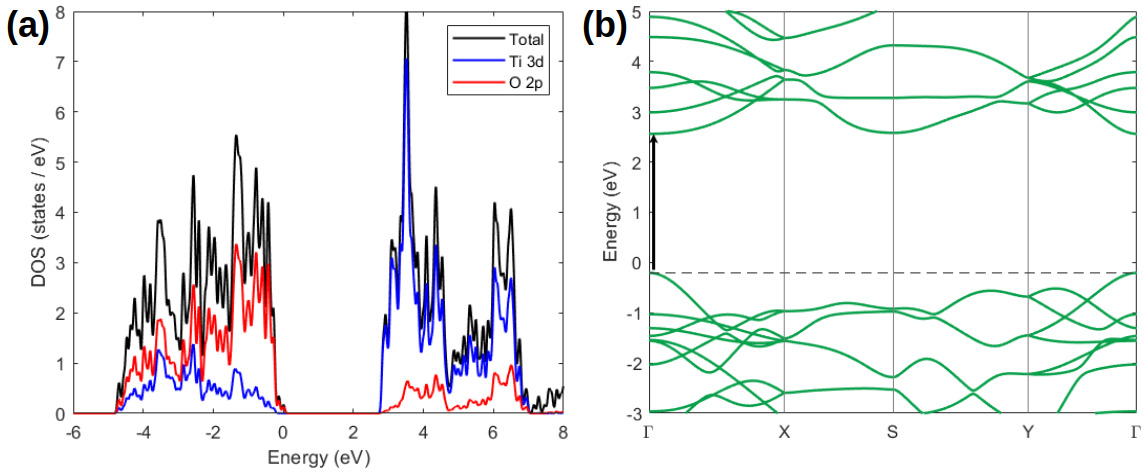}
\caption{(a) Density of states and (b) band structure of \ch{TiO2} monolayer with the GGA. The predicted band gap was found to be 2.76 eV, and is indicated with a black line in (b).}\label{DOS}
\end{figure}

\begin{figure}[h!]\centering
\includegraphics[width=1\linewidth]{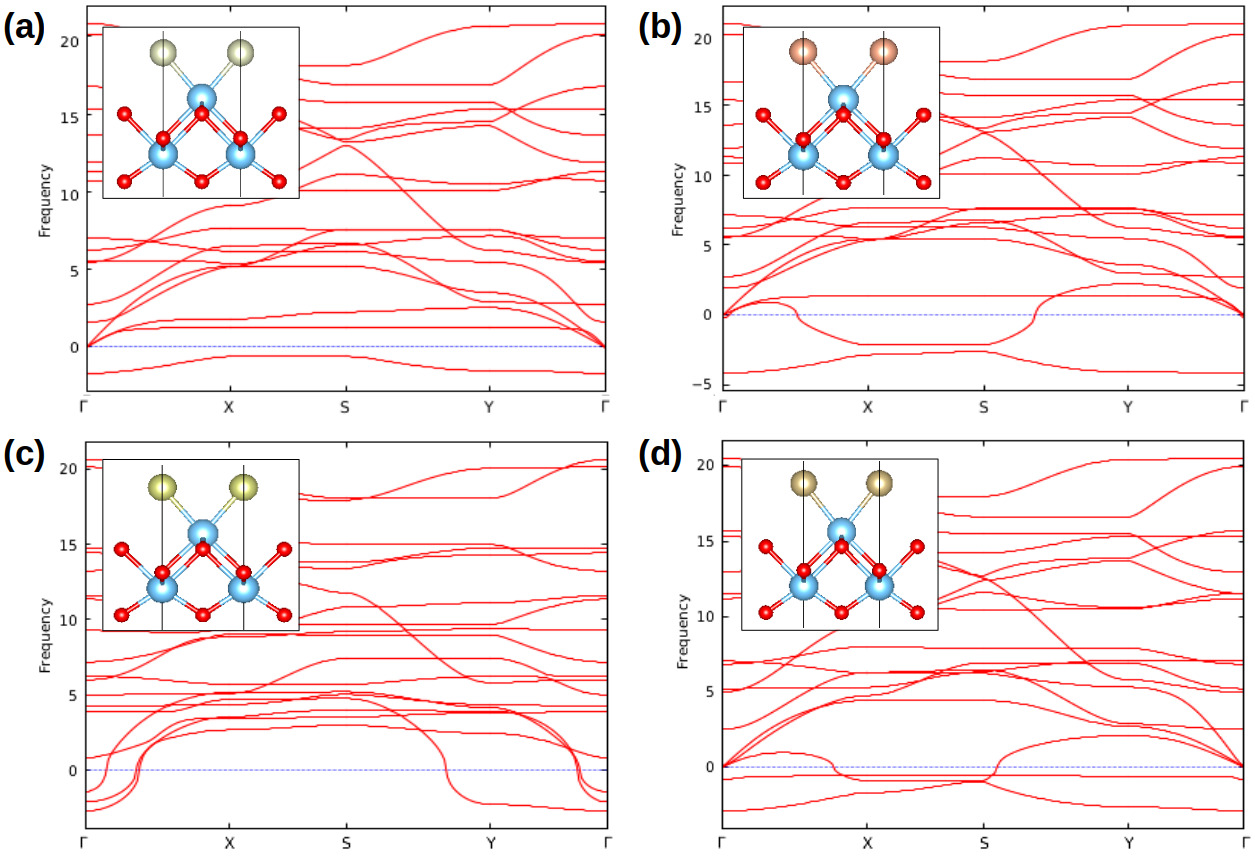}
\caption{Phonon band structure of (a) Rh-doped, (b) Ru-doped, (c) Ir-doped and (d) Os-doped \ch{TiO2} monolayers in which a bridging oxygen atom $O_b$ was replaced by noble metal atom. There exist imaginary phonon frequencies in the structures which indicates dynamical instability.}
\end{figure}

\begin{figure}[h!]\centering
\includegraphics[width=1\linewidth]{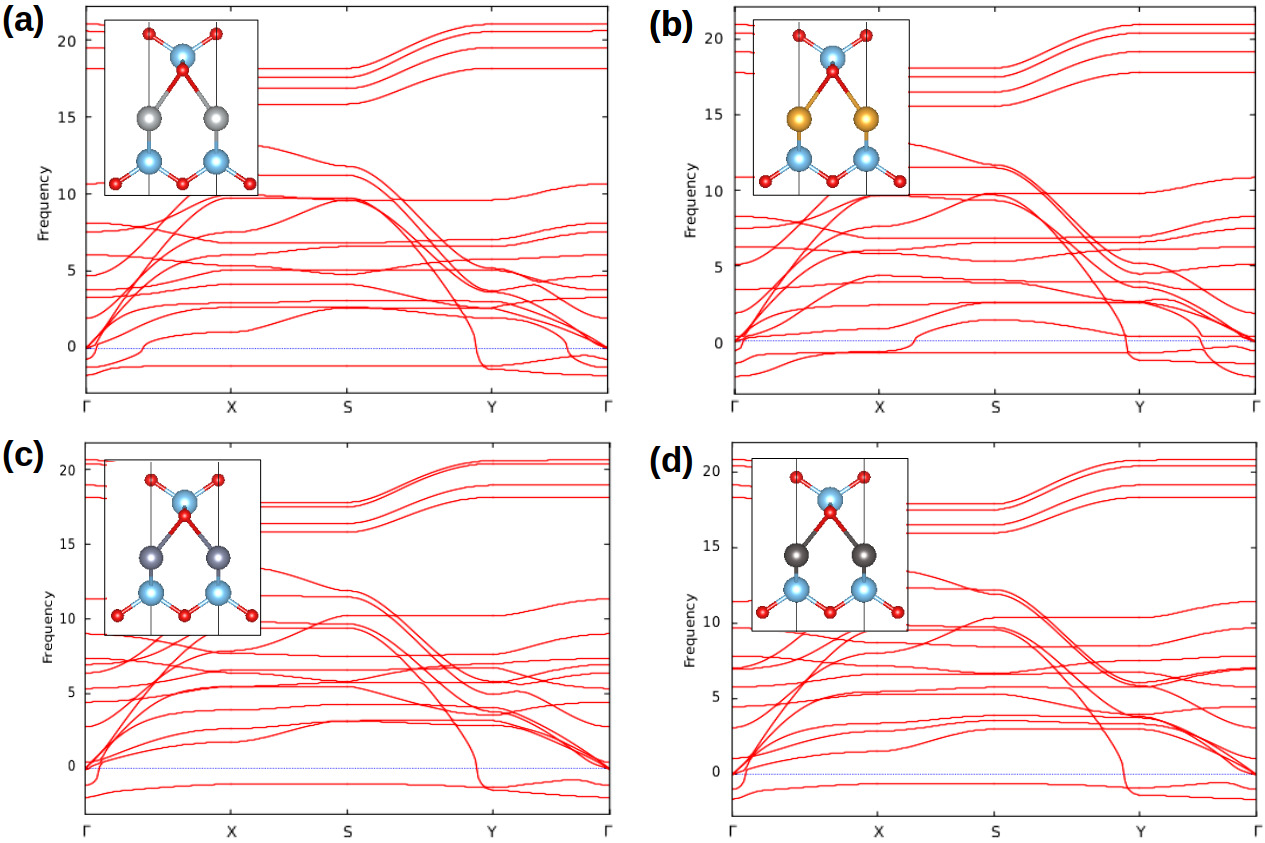}
\includegraphics[width=1\linewidth]{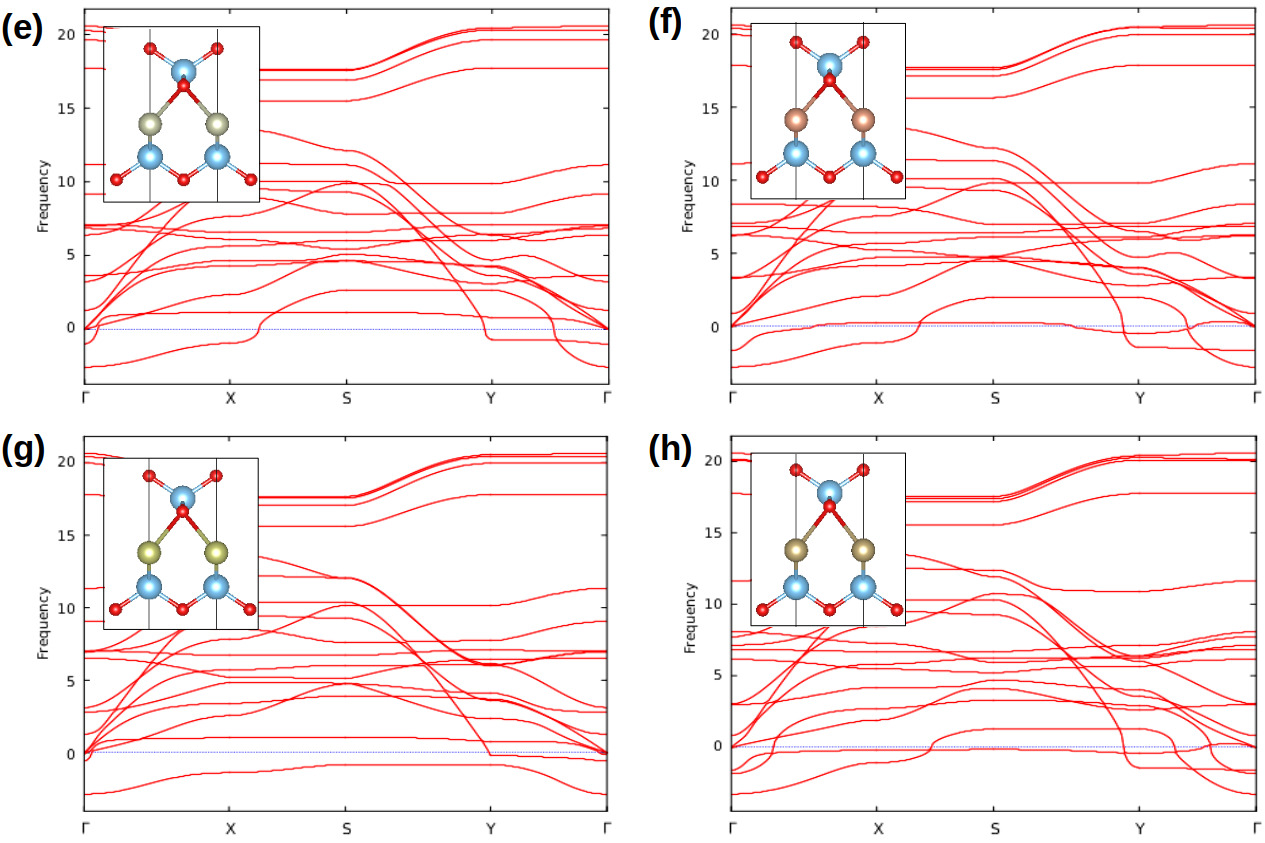}
\caption{Phonon band structure of (a) Ag-doped, (b) Au-doped, (c) Pd-doped, (d) Pt-doped, (e) Rh-doped, (f) Ru-doped, (g) Ir-doped and (h) Os-doped \ch{TiO2} monolayers in which an in-layer oxygen atom $O_i$ was replaced by noble metal atom. The presence of imaginary phonon modes shows instability for each system.}
\end{figure}

\begin{figure}[h!]\centering
\includegraphics[width=1\linewidth]{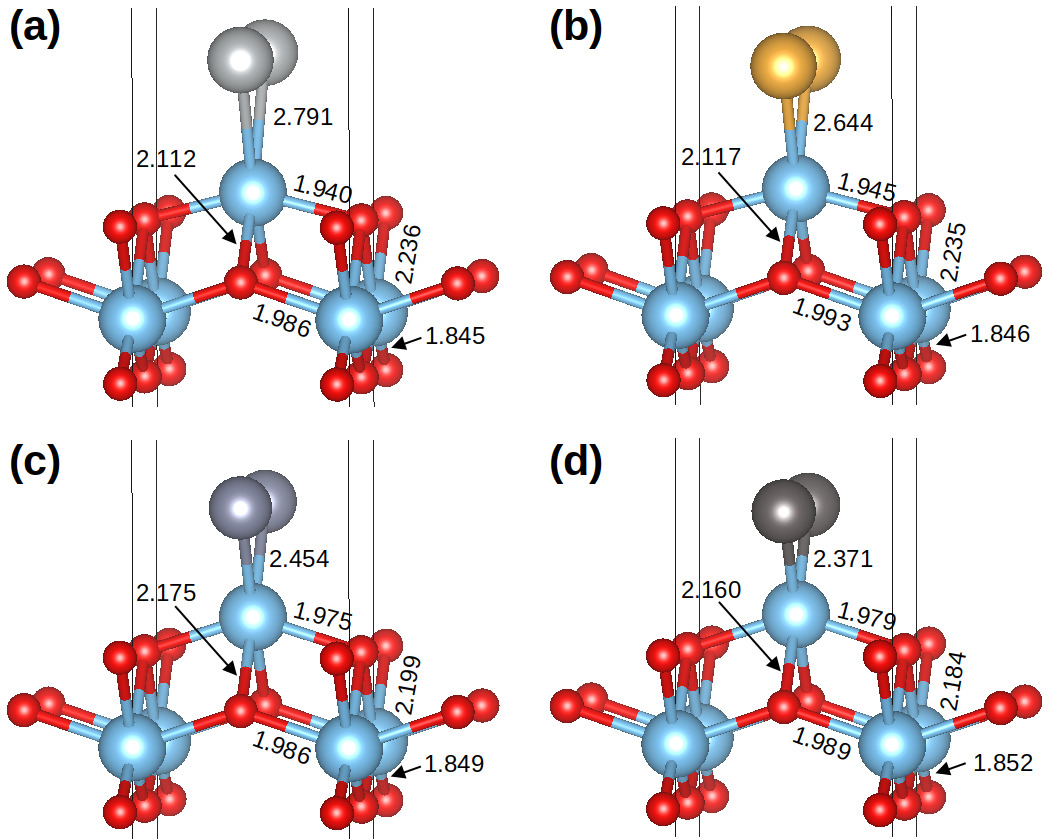}
\caption{Ti-O and Ti-noble metal distances in the (a) Ag-doped, (b) Au-doped, (c) Pd-doped and (d) Pt-doped \ch{TiO2} monolayers. Bond lengths are given in \AA ngstr{\"o}m (\AA).}
\end{figure}

\clearpage

\begin{figure}[h!]\centering
\includegraphics[width=1\linewidth]{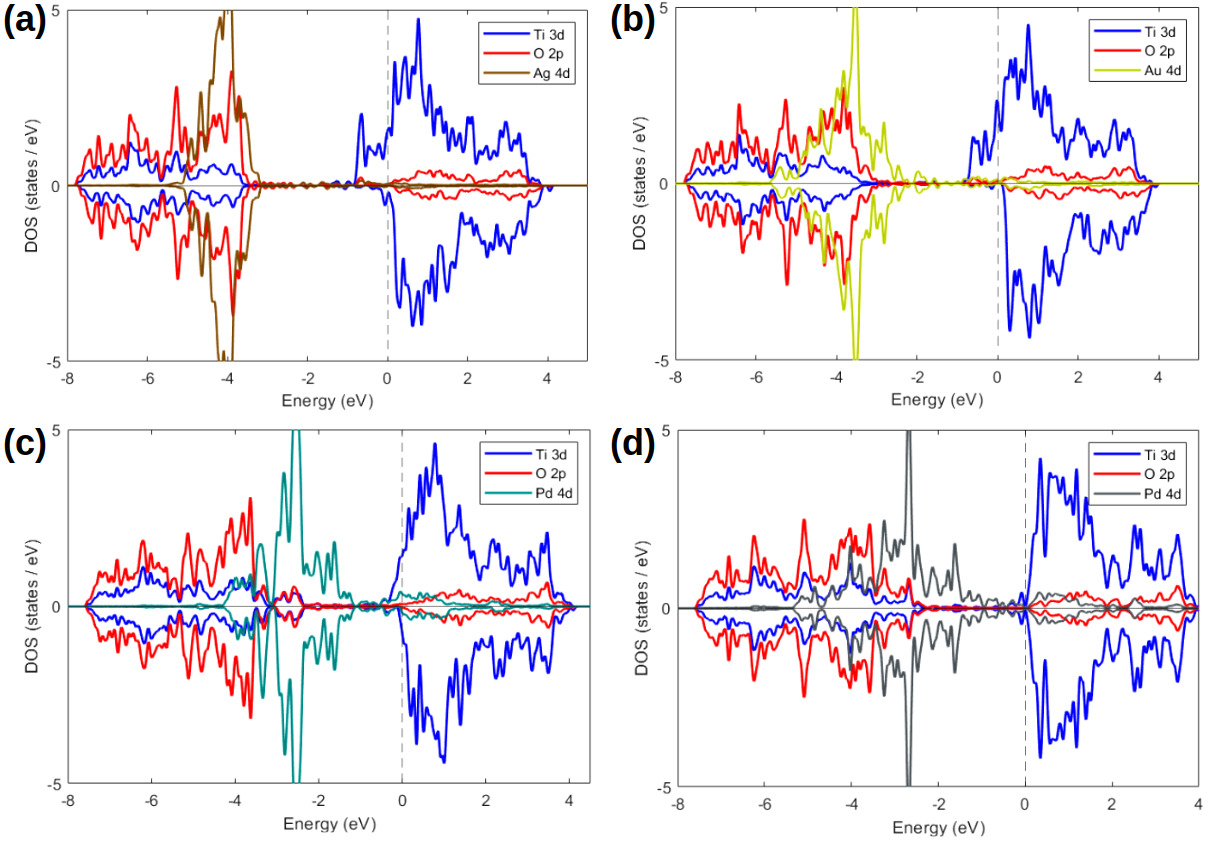}
\caption{Density of states of (a) Ag-doped, (b) Au-doped, (c) Pd-doped and (d) Pt-doped \ch{TiO2} monolayers using the GGA functional. The Fermi level is set to zero and indicated with black dashed line in the graphs.}\label{DOS}
\end{figure}

\begin{figure}[h!]\centering
\includegraphics[width=1\linewidth]{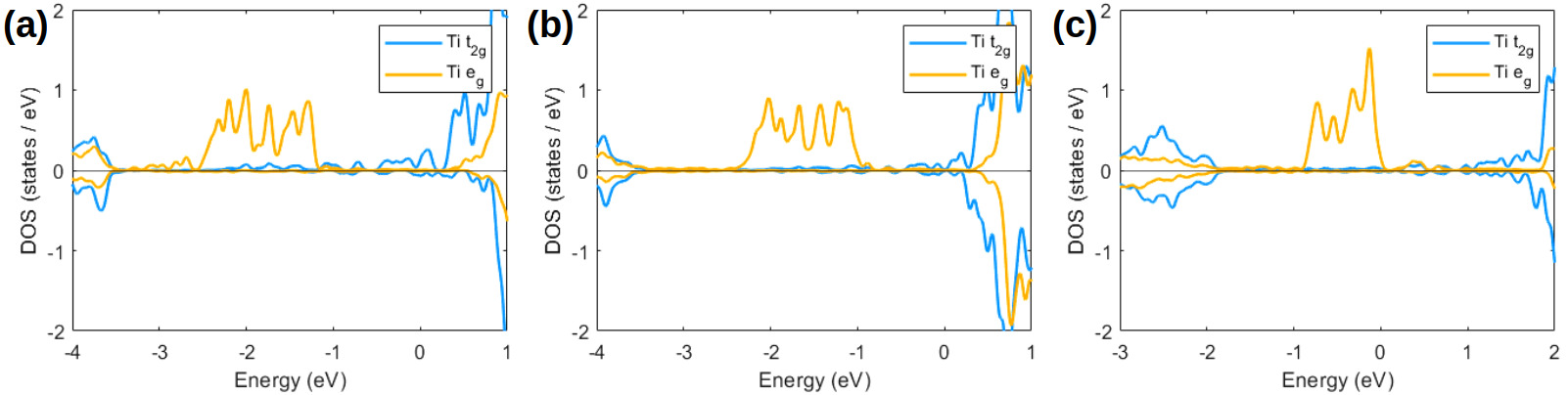}
\caption{Density of states of (a) Ag-doped, (b) Au-doped, (c) Pd-doped and (d) Pt-doped \ch{TiO2} monolayers using the GGA functional. The Fermi level is set to zero and indicated with black dashed line in the graphs.}\label{DOS}
\end{figure}

\begin{figure}[h!]\centering
\includegraphics[width=1\linewidth]{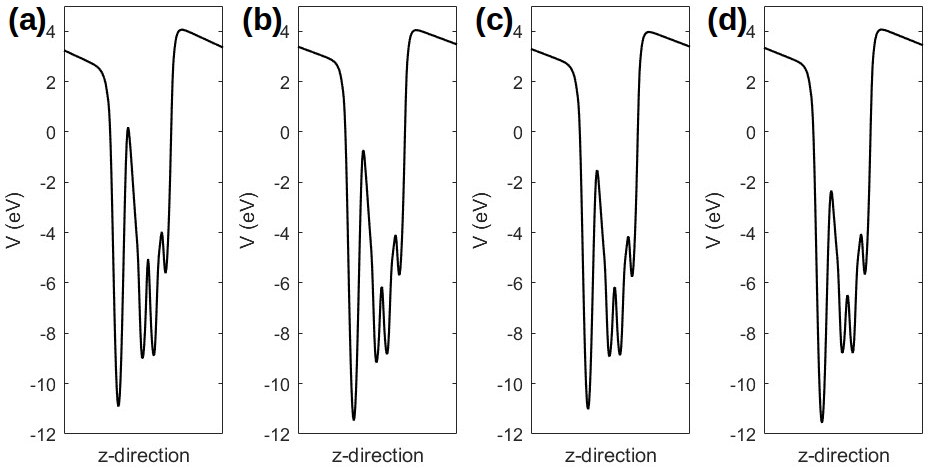}
\caption{Electrostatic potential of (a) Ag-doped, (b) Au-doped, (c) Pd-doped and (d) Pt-doped \ch{TiO2} monolayers without the dipole correction. In the vacuum the potential is not flat, and the work function cannot be determined accurately.}\label{WF}
\end{figure}

\begin{table}[h]\centering
\begin{tabular}{llclclc}
\textbf{System} &  & \multicolumn{1}{l}{\textbf{Ag-\ch{TiO2}}} &  & \multicolumn{1}{l}{\textbf{Au-\ch{TiO2}}} &  & \multicolumn{1}{l}{\textbf{Pd-\ch{TiO2}}} \\ \hline 
\rule{0pt}{4ex}Ti1                  &  & 0.047                      &  & 0.168                       &  & 0.005                       \\
\rule{0pt}{4ex}Ti2                  &  & 0.898                       &  & 0.778                        &  & 0.231                       \\
\rule{0pt}{4ex}O1                   &  & -0.026                      &  & -0.025                      &  & -0.005                      \\
\rule{0pt}{4ex}O2                   &  & 0.007                      &  & 0.002                      &  & 0.006                       \\
\rule{0pt}{4ex}O3                   &  & -0.004                     &  &  0.015                     &  & 0.00                       \\
\rule{0pt}{4ex}X                   &  & 0.015                       &  & 0.025                       &  & 0.018                       \\ 
\rule{0pt}{4ex}Total                &  & 0.937                        &  & 0.932                       &  & 0.255   \\[0.2cm] \hline                  
\end{tabular}
\caption{Magnetic moments (in $\mathrm{\mu_B}$) in the Ag-\ch{TiO2}, Au-\ch{TiO2} and Pd-\ch{TiO2} with the GGA. X specifies the noble metal atom (X$=$Ag, Au, Pd). Total magnetic moment of the unit cells, corresponding to the sum of local magnetic moments of atoms, is also provided.}
\end{table}
\clearpage

\begin{figure}[h!]\centering
\includegraphics[width=0.8\linewidth]{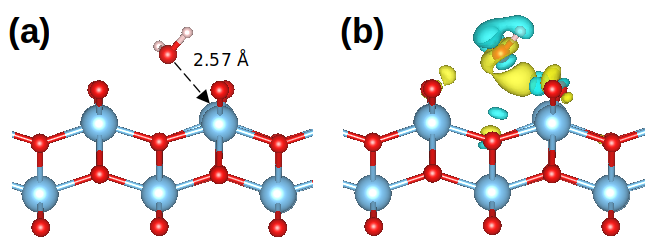}
\caption{(a) Water adsorption on the pristine \ch{TiO2} monolayer. The corresponding charge density difference plot is shown in (b). Isosurface value is set to 0.002 \textit{e}/\AA$^3$.}\label{WF}
\end{figure}

\begin{figure}[h!]\centering
\includegraphics[width=0.8\linewidth]{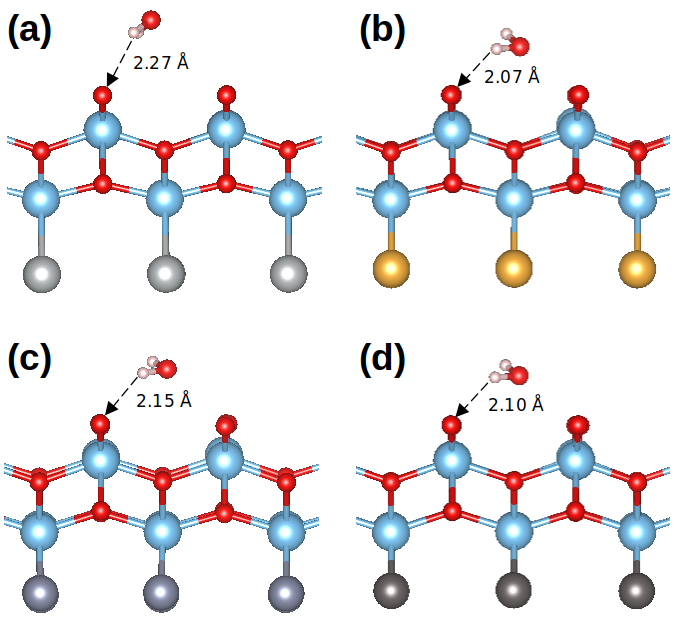}
\caption{The most stable adsorption configuration of water on the oxygen surface of the Janus monolayers: (a) Ag-\ch{TiO2}, (b) Au-\ch{TiO2}, (c) Pd-\ch{TiO2} and (d) Pt-\ch{TiO2}. The shortest distance between \ch{H2O} and a surface oxygen atom is given in \AA.}\label{WF}
\end{figure}

\begin{figure}[h!]\centering
\includegraphics[width=0.8\linewidth]{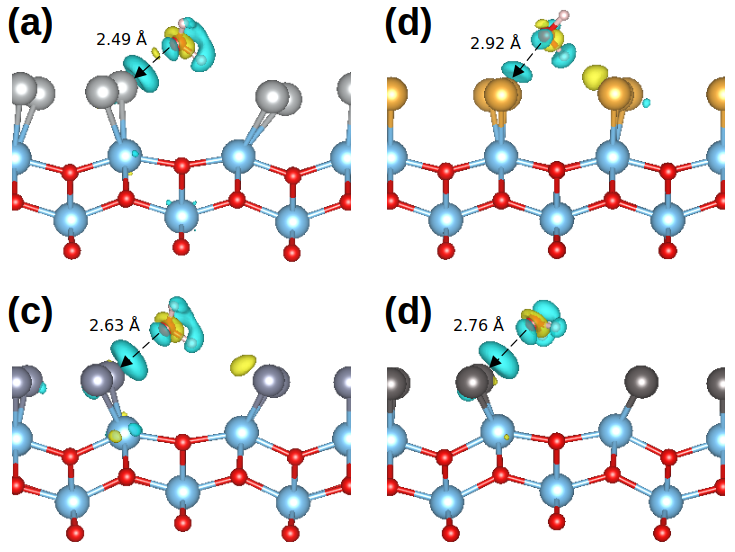}
\caption{The most stable adsorption configuration of water on the Janus monolayers with the corresponding charge density difference: (a) Ag-\ch{TiO2}, (b) Au-\ch{TiO2}, (c) Pd-\ch{TiO2} and (d) Pt-\ch{TiO2}. The shortest distance between \ch{H2O} and a noble metal atom is given in \AA, and isosurfaces are plotted at 0.002 \textit{e}/\AA$^3$.}\label{WF}
\end{figure}

\begin{figure}[h!]\centering
\includegraphics[width=1\linewidth]{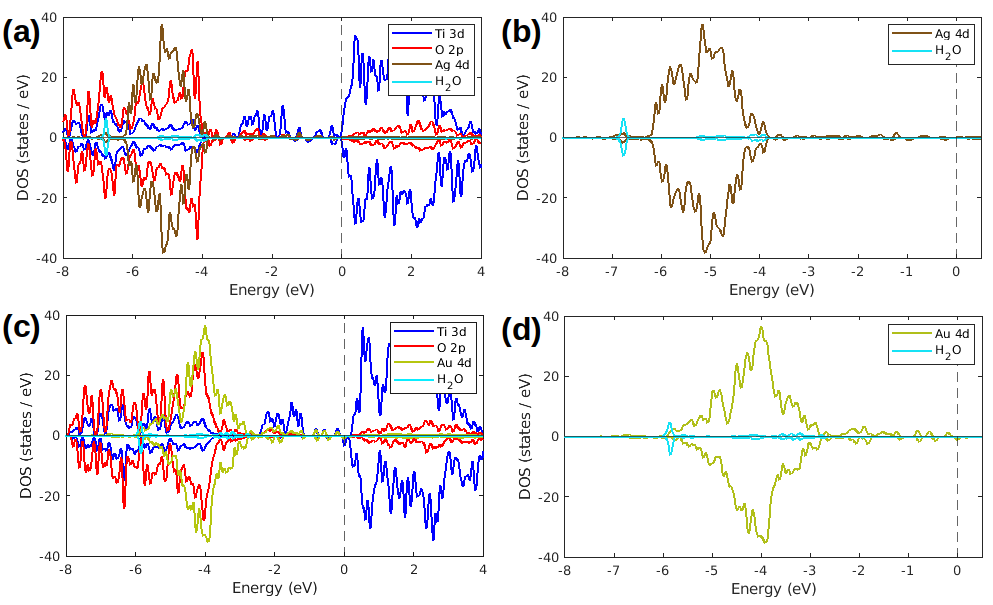}
\includegraphics[width=1\linewidth]{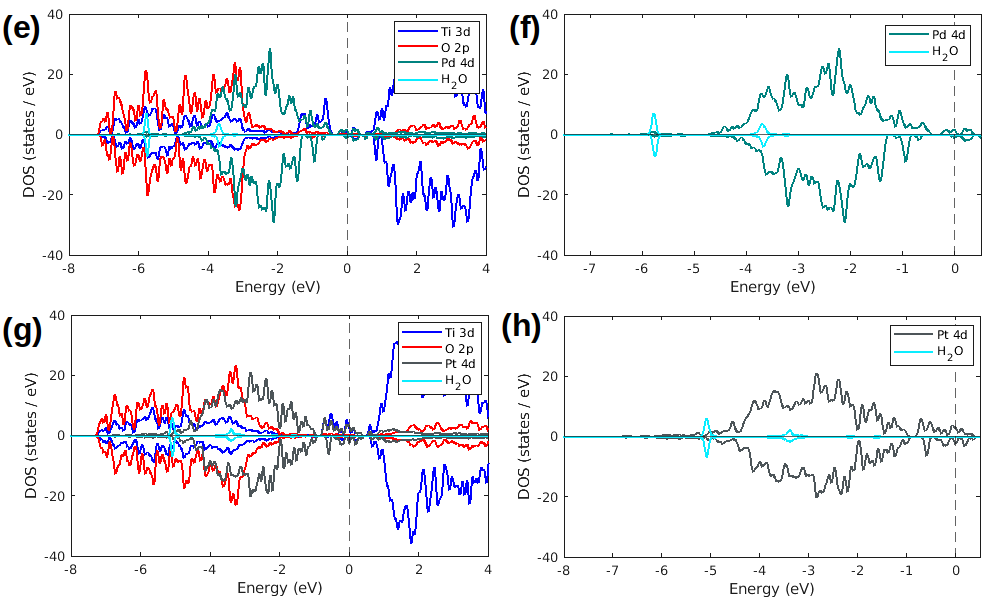}
\caption{Density of states of (a) Ag-doped, (c) Au-doped, (e) Pd-doped and (f) Pt-doped \ch{TiO2} monolayers with water adsorption on the noble metal surface using the GGA+U functional. On the right column (b, d, f and h) only partial density of states of noble metal and \ch{H2O} states are shown. The Fermi level is set to zero and indicated with black dashed line in the graphs.}
\end{figure}

\begin{table}[]
\scalebox{0.91}{
\begin{tabular}{llllllll}
        & \multicolumn{3}{l}{\quad\quad\quad\quad \textbf{At O surface}} & \multicolumn{4}{l}{\quad\quad\quad \textbf{At noble metal surface}} \\ \cline{2-8} 
\rule{0pt}{4ex}  & \textbf{H-O-H angle (\textdegree)}   &   & \textbf{d(H-O) (\AA)}    &  & \textbf{H-O-H angle (\textdegree)}      &    & \textbf{d(H-O) (\AA)} \\ \hline
\rule{0pt}{4ex}\textbf{Ag-TiO2} & \quad 103.32 &   &  0.971, 0.974  &   & \quad 105.12  &    &  0.976, 0.980   \\
\rule{0pt}{4ex}\textbf{Au-TiO2} & \quad 105.39 &   &  0.971, 0.978  &   & \quad 104.76  &    &  0.975, 0.982   \\
\rule{0pt}{4ex}\textbf{Pd-TiO2} & \quad 105.00 &   &  0.971, 0.978  &   & \quad 104.71  &    &   0.974, 0.984  \\
\rule{0pt}{4ex}\textbf{Pt-TiO2} & \quad 104.68 &   &  0.977, 0.980  &   & \quad 104.05  &    &  0.977, 0.977   \\[0.2cm] \hline 
\end{tabular}}
\caption{H-O-H angle and H-O bond lengths in the adsorbed water molecule on the Janus monolayers. For an isolated \ch{H2O} molecule, the calculated H-O-H angle was 104.60\textdegree and the H-O bond length was 0.973 \AA, being in line with experimentally measured values \cite{water}. On the surface of pristine \ch{TiO2} monolayer, the H-O bond lengths were 0.973 and 0.975 \AA\ and H-O-H angle was 106.84\textdegree.}
\end{table}
\clearpage

\subsection*{Calculation of the Gibbs free energy of the HER reaction}

The Gibbs free energy of hydrogen adsorption was calculated as in N{\o}rskov scheme \cite{Norskov}. In general, the vibrational frequency of hydrogen can be negligible, and therefore it was approximated from the equation $\Delta S \approx -\frac{1}{2}S(\ch{H2})$ where $S(\ch{H2})$ is the entropy of the \ch{H2} molecule. Besides, the equation $\Delta E_{\mathrm{ZPE}}=E_{\mathrm{ZPE}}(H^*)-\frac{1}{2}E_{\mathrm{ZPE}}(\ch{H2})$ was used to estimate the zero-point energy of adsorbed hydrogen. The zero-point energy of adsorbed hydrogen $E_{\mathrm{ZPE}}(H*)$ was calculated from the vibrational frequency of H atoms at $298.15 K$. Both zero-point energy $E_{\mathrm{ZPE}}(\ch{H2})$ and entropy $S(\ch{H2})$ of the \ch{H2} molecule were obtained from the NIST-CCCBDB database \cite{nist} at standard conditions ($T=298.15 K$). Thus, considering the $E_{\mathrm{ZPE}}$ and $\Delta S$, the expression for Gibbs free energy,
\begin{equation}
\Delta G = E_{\mathrm{ads}}+E_{\mathrm{ZPE}}-T\Delta S
\end{equation}
can be rewritten as
\begin{equation}
\Delta G = E_{\mathrm{ads}}+0.2002
\end{equation}
The correction term is close to the value reported by N{\o}rskov \textit{et al.} (0.24 eV) \cite{Norskov}.

\begin{figure}[h!]\centering
\includegraphics[width=0.75\linewidth]{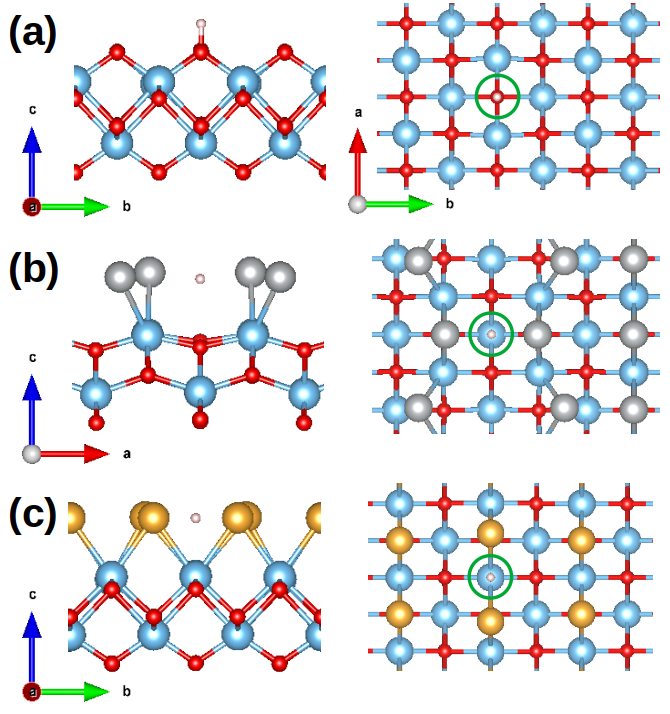}
\includegraphics[width=0.75\linewidth]{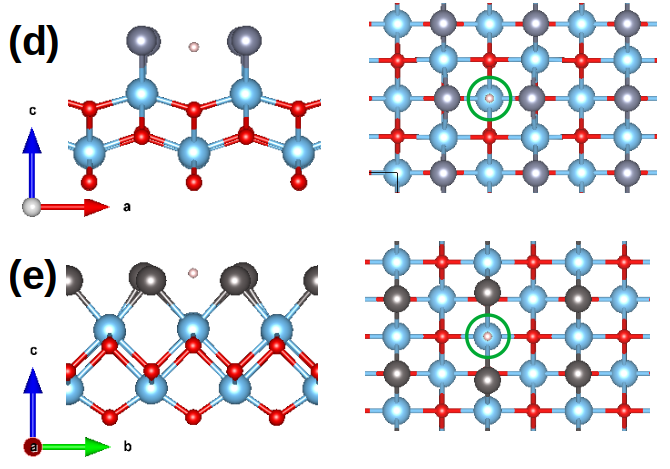}
\caption{Side (either along x or y-direction) and top view of the most stable adsorption structure of hydrogen on (a) \ch{TiO2}, (b) Ag-\ch{TiO2}, (c) Au-\ch{TiO2}, (d) Pd-\ch{TiO2} and (e) Pt-\ch{TiO2}. The hydrogen atom is indicated as white, and the adsorption site is highlighted with a green circle.}
\end{figure}

%\begin{table}[h]\centering
%\begin{tabular}{llclclclc}
%\textbf{System} &  & \multicolumn{1}{l}{\textbf{Ag-\ch{TiO2}}} &  & \multicolumn{1}{l}{\textbf{Au-\ch{TiO2}}} &  & \multicolumn{1}{l}{\textbf{Pd-\ch{TiO2}}} &  & \multicolumn{1}{l}{\textbf{Pt-\ch{TiO2}}} \\ \hline 
%\rule{0pt}{4ex}Ti1                  &  & 1.89                       &  & 1.88                       &  & 1.92  & & 1.92                      \\
%\rule{0pt}{4ex}Ti2                  &  & 1.54                       &  & 1.63                        &  & 1.57 & & 1.64                       \\
%\rule{0pt}{4ex}O1                   &  & -1.10                      &  & -1.09                      &  & -1.09 & & -1.08                     \\
%\rule{0pt}{4ex}O2                   &  & -1.08                      &  & -1.08                      &  & -1.06 & & -1.07                      \\
%\rule{0pt}{4ex}O3                   &  & -0.877                      &  & -0.876                      &  & -0.867 & & -0.874                      \\
%\rule{0pt}{4ex}NM                   &  & -0.367                       &  & -0.468                       &  & -0.471 & &  -0.540    \\[0.2cm] \hline                  
%\end{tabular}
%\caption{Bader charges (in \textit{e}) in the Ag-\ch{TiO2}, Au-\ch{TiO2}, Pd-\ch{TiO2} and Pt-\ch{TiO2} with the GGA.}
%\end{table}